# Autoignition of two-phase *n*-heptane/air mixtures behind an oblique shock: insights into spray oblique detonation initiation

Hongbo Guo[1,2], Yong Xu[2], Ningbo Zhao[1], Huangwei Zhang[2,*]

*[1] College of Power and Energy Engineering, Harbin Engineering University*
*Harbin 150001, People's Republic of China*
*[2] Department of Mechanical Engineering, National University of Singapore*
*9 Engineering Drive 1, Singapore, 117576, Republic of Singapore*

**Abstract**

Autoignition of *n*-heptane droplet/vapor/air mixtures behind an oblique shock wave are studied, through Eulerian-Lagrangian method and a skeletal chemical mechanism. The effects of gas/liquid equivalence ratio (ER), droplet diameter, flight altitude, and Mach number on the ignition transient and chemical timescales are investigated. The results show that the ratio of chemical excitation time to ignition delay time can be used to predict the oblique detonation wave (ODW) transition mode. When the ratio is relatively high, the combustion heat release is slow and smooth transition is more likely to occur. In heterogeneous ignition, there are direct interactions between the evaporating droplets and the induction/ignition process, and the chemical explosive propensity changes accordingly. The energy absorption of evaporating droplets significantly retards the ignition of *n*-heptane vapor. In the two-phase *n*-heptane mixture autoignition process, the ignition delay time decreases exponentially with flight Mach number, and increases first and then decreases with the flight altitude. As the liquid ER increases, both ignition delay time and droplet evaporation time increase. With increased droplet diameter, the ignition delay time decreases, and the evaporation time increases. Besides, for Mach number is less than 10, the ratio of the chemical excitation time to ignition delay time generally increases with the flight altitude or Mach number. It increases when the liquid ER decreases or droplet diameter increases. When Mach number is sufficiently high, it shows limited change with fuel and inflow conditions. The results from this work can provide insights into spray ODW initiation. The ODW is more likely to be initiated with a smooth transition at high altitude or Mach number. Abrupt transition mode tends to happen when fine fuel droplets are loaded.

**Keywords:** *n*-Heptane; Droplet evaporation; Autoignition; Oblique detonation; Transition mode; Spray detonation

---

*Corresponding author. Tel.: +65 6516 2557; Fax: +65 6779 1459.
*E-mail address*: huangwei.zhang@nus.edu.sg.

# 1. Introduction

Detonation has been recognized as a promising combustion mode. Its thermal cycle efficiency is higher than that of a constant-pressure combustion, and is close to that of a constant-volume combustion [1,2]. In recent years, to develop novel propulsion technology, detonation engines have attracted increased interests from both scientific and engineering communities. Oblique Detonation Engine (ODE) is deemed a new propulsion device for air-breathing hypersonic aircraft [3]. It not only has the advantages of scramjet (e.g., simpler combustion chamber and wider flight Mach numbers), but can also generate steady thrust [4,5]. In an ODE, oblique detonation wave (ODW) can stand over a finitely long wedge due to the coupling between an oblique shock wave (OSW) and reactions of supersonic combustible [6].

Initiation and structure of oblique detonation waves are first studied by Li et al. through numerical simulations [6]. Their work reveals that the ODW structure is usually composed of a non-reactive OSW, an induction zone, deflagration wave, and an oblique detonation wave. Two transition modes between OSW and ODW are observed: smooth mode featured by a curved shock, while abrupt mode with a multi-wave point. Subsequent experiments and simulations further confirm the existence of these modes [7,8]. Considering different inflow parameters (e.g., flight altitude and Mach number), Teng et al. [9] identify four structures in the transition zone. Meanwhile, the morphology of the ODWs is predicted by the geometric analysis of double Mach lines and the relation of Mach numbers. Previous results also demonstrate that transition mode strongly depends on the inflow Mach number, but is less effected by the inflow pressure [10]. The abrupt transition would result in formation of a primary transverse wave in the initiation zone [11,12]. Guo et al. [13] explore the predictability of the shock-detonation transition mode of ODW in *n*-heptane/air through the relation between ignition delay and chemical excitation time. In spite of the foregoing insightful studies, what are the controlling factors for the transition mode from shock to detonation has not reached a consensus.

Most previous ODW studies consider gaseous fuels (e.g. hydrogen, methane or acetylene) [14–16]. However, for practical applications, liquid fuels are preferred because of small storage space and high energy density. Kailasanath [1] summarizes the main development and applications on liquid-fueled detonation propulsions before 2000. In this century, the interests in liquid-fueled detonation have revived and considerable tests have been reported. For instance, Fan et al. [17] successfully perform two-phase pulse detonation experiments with liquid $C_8H_{16}$/air mixture, and find that use of liquid fuels can reduce the combustor size and increases the frequency of power output. Bykovskii et al. [18] investigate the propagation and propulsive performance of Rotating Detonation Engine's (RDE's) using liquid kerosene with hydrogen or syngas. Recently, Wolański et al. [19] also successfully perform RDE tests with partially mixed preheated liquid Jet-A and air.

As for ODE, Ren et al. [20,21] numerically study the effects of liquid fuel equivalence ratio on ODW initiation and the OSW-ODW transition in two-phase kerosene-air mixtures. They find that the variations of initiation length are dominated by evaporative cooling, and the transition structure shifts from a smooth mode to an abrupt one when the liquid fuel flow rate increases. Guo et al. [22] simulate the hydrogen ODW's in water mist flows, and analyze the chemical and physical effects of liquid droplets on the induction zone chemistry and ODW initiation. However, how the liquid fuel droplets modulate the post-OSW chemistry and hence the detonation transition mode have not been studied yet.

The objective of this paper is to analyze how the mass and heat transfer caused by droplet evaporation affect the autoignition under ODW-relevant conditions. *n*-Heptane is considered as the fuel, which is a good representation for numerous realistic hydrocarbon fuels. Similar to the work by Bouali et al. [23], a constant-volume reactor is used to study the autoignition of heterogeneous mixtures containing *n*-heptane droplets, vapor, and air. The initial pressure and temperature of the reactor are determined based on the flight conditions (i.e., altitude and Mach number). The droplet phase is simulated with the Lagrangian approach and particle-source-in-cell method [24] is adopted to model the droplet effects on the gas phase. A skeletal mechanism (44 species and 112 elementary reactions [25]) is employed for *n*-heptane combustion. The influences of droplet equivalence ratio, diameter and background gas equivalence ratio will be examined through parametric studies. The structure of the manuscript is organized as below. The physical problem is detailed in Section 2, whereas the ODW-relevant chemical timescales and transition mode are introduced in Sections 3 and 4, respectively. The results will be discussed in Section 5, followed by conclusions in Section 6.

## 2. Problem description

### 2.1 Background

The schematic of an ODW engine is shown in Fig. 1, following Dudebout et al. [26], which is composed of three sections, i.e., inlet, combustor, and nozzle. The supersonic flow is compressed by two attached shocks induced by the inlet with the same relative deflection angle $\delta$ to minimize entropy production [27]. Under the designed flight condition, they would intersect at the tip of the cowl (C in Fig. 1a). The fuel is injected into the inlet flow parallel to the oncoming air flows ($O_2$: $N_2$ = 1: 3.76 by vol.). Reactant mixing would be facilitated due to relatively long residence time of the fuel/air through the forebody of the vehicle. However, perfect reactant mixing before the combustor is still challenging. In our study, it is assumed that the fuel and air are premixed at the combustor entrance (BC in Fig 1a).

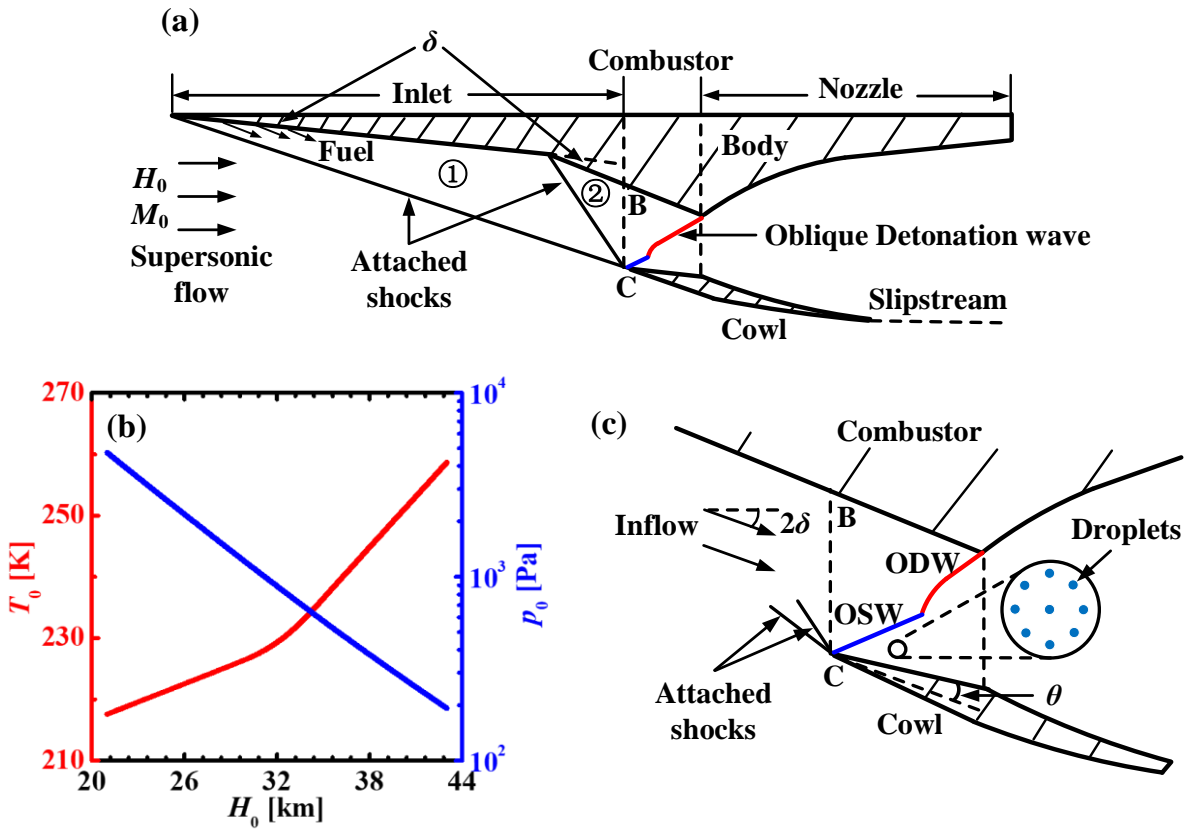

Figure 1: (a) Schematic of an ODW engine, (b) standard atmosphere parameters with different flight altitudes, and (c) zoom-in view of the combustor [26,28,29].

Typically, flight conditions can be parameterized by the Mach number $M_0$ and flight altitude $H_0$.

The freestream velocity $U_0$ can be obtained from $M_0$. For instance, at $H_0 = 25$ km, $U_0 = 3017.09$ m/s for $M_0 = 10$ and $U_0 = 3620.51$ m/s for $M_0 = 12$. Moreover, according to the standard atmosphere [29], the pressure and temperature of the freestream supersonic flows, $p_0$ and $T_0$, can be determined by the flight altitude $H_0$. Air-breathing vehicles equipped with ODE normally operate at high altitudes of 25 – 40 km [9,30]. Within this range, $p_0$ and $T_0$ respectively increases and decreases with $H_0$, see Fig. 1(b). For example, $T_0 = 221.55$ K and $p_0 = 2549.10$ Pa at $H_0 = 25$ km, while $T_0 = 236.51$ K and $p_0 = 574.56$ Pa at $H_0 = 35$ km.

The oncoming flows are compressed twice by two attached shocks. The thermodynamic parameters after each shock can be calculated through the Rankine-Hugoniot relations for a given attached shock wave angle $\beta_i$ ($i = 1, 2$). They correspond to states ① and ② in Fig. 1(a). We can obtain the pressure and temperature ratios before and after the attached shock [31]

$$\frac{p_b}{p_f} = 1 + \gamma_f M_{fn}^2(1 - X), \tag{1}$$

$$\frac{T_b}{T_f} = \frac{1}{[\gamma_b(\gamma_f-1)]/[\gamma_f(\gamma_b-1)]} \frac{R_{gf}}{R_{gb}} \left[1 + \frac{\gamma_f-1}{2} M_{fn}^2(1 - X^2)\right], \tag{2}$$

where the subscripts $f$ and $b$ represent the parameters before and after the shock, respectively. The subscript $n$ represents the parameters normal to the shock wave. $\gamma$ is the specific heat ratio, $R_g$ is the specific gas constant, and $M_{fn} = M_f \sin\beta$ is the normal Mach number of the shock. The density ratio across the shock is $X = \rho_f/\rho_b = \tan(\beta - \delta)/\tan\beta$.

By iteratively solving Eq. (1) and (2), the flow parameters after two attached shocks, i.e. $p_2$, $T_2$ and $U_2$ in zone ②, can be determined by a given flow deflection angle $\delta$ (see the procedures in [32]). These parameters are taken as the inflow conditions of the combustor section, starting from line BC in Fig. 1(a) or a zoom-in view in Fig. 1(c). The premixed gas enters the combustor and then reflects on the wedge with deflection angle $\theta$, generating an oblique shock wave. The oblique shock wave may trigger chemical reactions and subsequently induce an oblique detonation wave downstream [6,8,33]; see Fig. 1(c). Finally, an oblique detonation is formed. The high-speed hot products leave the nozzle to generate thrust.

## 2.2 Physical model

Ignition of supersonic fuel/air mixtures behind an oblique shock wave induced by a semi-infinite wedge under high-altitude flight conditions will be studied. It is known that the ignition process behind an oblique shock wave (i.e., ODW induction zone) is approximately constant-volume combustion. A closed reactor is therefore used, as illustrated in Fig. 2. It mimics a control volume $V_0$ behind the oblique shock wave, containing a heterogeneous mixture of *n*-heptane droplet/vapor/air mixtures, as shown in Fig. 1(c). The size of the reactor is $1\times1\times1$ mm$^3$, which is discretized with one CFD cell. This essentially leads to zero-dimensional (0D) calculations of ignition of two-phase mixtures. Similar method is also used by Bouali et al. [23], to study the effects of evaporating fuel droplets on *n*-heptane ignition transients.

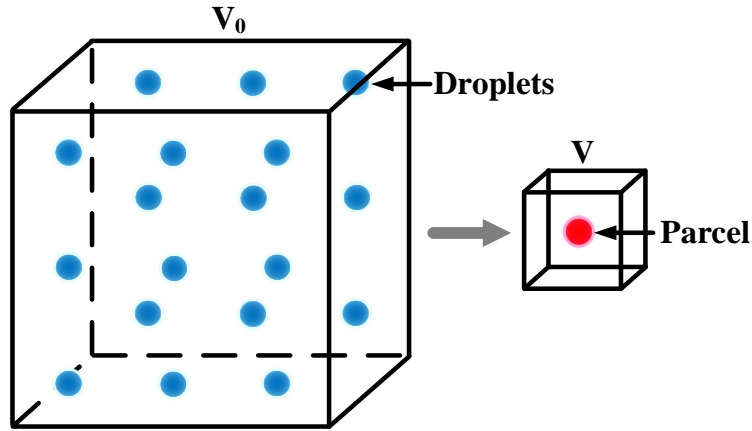

Figure 2: Schematic of a closed reactor with *n*-heptane droplet/vapor/air mixture.

The oncoming mixtures consist of *n*-heptane vapor, droplets, and air. The initial pressure and temperature in the reactor are calculated from the post-OSW thermodynamic state derived from $H_0$, $M_0$, $\theta$ and $\delta$ (see Figs. 1a and 1c). The flight altitude $H_0$ varies from 21 to 43 km, and Mach number $M_0$ varies from 8 to 15. In this study, the wedge angle and deflection angle are fixed to be $\theta = 20°$ and $\delta = 12.5°$, respectively. Although their variation may affect the ignition of two-phase mixtures, nonetheless, this is beyond the scope of this work and will be studied in a separate manuscript. By changing the droplet diameter, droplet number, initial thermo- and aero-dynamic parameters, autoignition of two-phase mixtures subject to different conditions can be examined.

Zero-gradient conditions are enforced for temperature, pressure and species mass fraction for six boundaries of the cubic domain. The equivalence ratio (ER) of the background premixed *n*-heptane vapor/air mixtures, $\phi_g$, ranges from 0.2 to 2.0. The resultant species molar ratio follows *n*-$C_7H_{16}/O_2/N_2 = \phi_g/11/41.36$. Existence of the fuel vapor mimics the possible droplet evaporation in the shocked gas (i.e., zone ① and ②) after the liquid fuel is atomized.

Ultrafine *n*-heptane droplets are distributed in the reactor, see Fig. 2, and the initial droplet diameter, $d_d^0$, are 5 to 50 μm. This range is deemed reasonable for ODE configuration, in light of the strong aerodynamic fragmentation effects by high-speed flows and shock waves. Due to relatively low freestream pressure and temperature (Fig. 1b), the droplets would vaporize in the combustor under subcritical conditions [34,35]. The initial material density, heat capacity, and temperature are 680 kg/m³, 2952 J/kg/K, and 298 K, respectively [36]. In our calculations, the liquid phase ER is $\phi_l = 0.0$–2.0, estimated based on the mass of liquid *n*-heptane and oncoming air.

## 2.3 Mathematical model

In this work, we only focus on heat and mass interactions due to droplet evaporation on autoignition of *n*-heptane droplet/vapor/air mixture. In the reactor, the background gas is stationary and gas motion caused by mass or heat transfer is neglected. Therefore, the gas velocity is always zero. Due to dilute sprays, the droplet volume effects are not taken into consideration. For gas phase, the equations of mass, energy, and species mass fraction are solved along with the equation of state, i.e., $p = \rho R_g T$. They respectively read

$$\frac{d\rho}{dt} = S_{mass}, \tag{3}$$

$$\frac{d(\rho E)}{dt} = \dot{\omega}_T + S_{energy}, \tag{4}$$

$$\frac{d(\rho Y_m)}{dt} = \dot{\omega}_m + S_{species,m}, (m = 1, \ldots M - 1). \tag{5}$$

In above equations, $t$ is time and $\rho$ is the gas density. $p$ is the gas pressure, $T$ is the gas temperature, and $R_g$ is specific gas constant. $E$ is the total non-chemical energy, in which $E = h_s - p/\rho$ is the sensible internal energy and $h_s$ is the sensible enthalpy [37]. $Y_m$ is the mass fraction of $m$-th species and $M$ is the total species number. The term $\dot{\omega}_T$ represents the heat release rate from chemical reactions, whilst $\dot{\omega}_m$ is the reaction rate of $m$-th species by all $N$ reactions.

The Lagrangian method is used to model the dispersed liquid phase, composed of a large number of spherical droplets. The temperature inside the droplet is assumed to be uniform, since the Biot number of the $n$-heptane droplet is small. The equations of mass and energy for a single droplet are

$$\frac{dm_d}{dt} = -\dot{m}_d, \tag{6}$$

$$c_{p,d}\frac{dT_d}{dt} = \frac{\dot{Q}_c + \dot{Q}_{lat}}{m_d}, \tag{7}$$

where $m_d = \pi\rho_d d_d^3/6$ is the mass of a single droplet, $\rho_d$ and $d_d$ are the droplet material density and diameter, respectively. $c_{p,d}$ is the droplet heat capacity, and $T_d$ is the droplet temperature.

The droplet evaporation rate $\dot{m}_d$ in Eq. (6) is modelled through $\dot{m}_d = \pi d_d Sh D_{ab}\rho_s \ln(1+X_r)$ [38]. The vapor molar ratio $X_r$ is estimated from $X_r = (X_S - X_C)/(1 - X_S)$, where $X_C$ is the vapor mole fraction in the surrounding gas, and $X_S$ is the vapor mole fraction at the droplet surface, calculated using Raoult's Law, i.e., $X_S = X_m p_{sat}/p$. $X_m$ is the mole fraction of the condensed species in the liquid phase. $p_{sat}$ is the saturation pressure and is a function of droplet temperature [36]. Moreover, $\rho_S = p_S W_m/RT_S$ is the vapor density at the droplet surface, $p_S$ is the surface vapor pressure, $T_s = (T + 2T_d)/3$ is the droplet surface temperature and $R = 8.314$ J/(mol·K) is the universal gas constant. $D_{ab}$ is the vapor diffusivity in the gaseous mixture, depending on droplet surface vapor pressure and temperature [39]. $Sh = 2.0 + 0.6Re_d^{1/2}Sc^{1/3}$ is the Sherwood number. In Eq. (7), $\dot{Q}_c = h_c A_d(T - T_d)$ is the convective heat transfer rate, in which $h_c$ is the convective heat transfer coefficient, computed by Ranz and Marshall correlation [40], i.e., $Nu = h_c d_d/k = 2.0 + 0.6Re_d^{1/2}Pr^{1/3}$. $Nu$ and $Pr$ are the Nusselt and Prandtl numbers of the gas phase, respectively, and $Re_d$ is the droplet Reynolds number. In addition, $\dot{Q}_{lat} = -\dot{m}_d h(T_d)$ is the heat transfer associated with droplet evaporation, $h(T_d)$ is the latent heat of vaporization at droplet temperature.

The influences of *n*-heptane droplets on the gas phase are modelled with Particle-source-in-cell (PSI-CELL) approach [24]. The terms, $S_{mass}$, $S_{energy}$ and $S_{species,m}$ in Eqs. (3) – (5), account for the mass, energy and species transfer between the gas and droplet phases. They are calculated from

$$S_{mass} = \frac{1}{V_c}\sum_{i=1}^{N_d} \dot{m}_{d,i}, \tag{8}$$

$$S_{energy} = -\frac{1}{V_c}\sum_{i=1}^{N_d}(-\dot{m}_{d,i}h_g + \dot{Q}_{c,i}), \tag{9}$$

$$S_{species,m} = \begin{cases} S_{mass} & for\ fuel\ species \\ 0 & for\ other\ species \end{cases}. \tag{10}$$

Here $V_c$ is the CFD cell volume, $N_d$ the droplet number in one cell, and $h_g$ the fuel vapor enthalpy at the droplet temperature.

The gas and liquid droplet equations are solved with a two-phase reacting flow solver, *RYrhoCentralFoam* [41,42]. It has been extensively validated and verified against the experimental or theoretical data in different problems, including autoignition and droplet evaporation [42]. It has been successfully applied for modelling various detonation and supersonic combustion problems [22,43–45]. In this solver, second-order implicit backward method is employed for temporal discretization and the time step is about $1\times10^{-9}$ – $1\times10^{-8}$ s. Chemistry integration is performed with a Euler implicit method, and its accuracy has been confirmed with other ordinary differential equation solvers in our recent work [13,42]. A skeletal mechanism with 44 species and 112 reactions [25] is used for *n*-heptane/air combustion. The accuracy of this mechanism has been compared with a more detailed chemistry (88 species and 387 reactions [46]), see Fig. S2 in supplementary document. The droplet equations, i.e., Eqs. (6) – (7), are solved by a first-order Euler method and the right-hand terms (e.g., $\dot{m}_d$ in Eq. 6) are integrated in a semi-implicit fashion. The gas properties at the droplet location are linearly interpolated from the gas phase results.

## 3. Ignition delay and chemical excitation time

In oblique detonations, constant-volume ignition of the mixture occurs behind the oblique shock [10,16]. The ignition delay time, $\tau_i$, is directly affected by the freestream condition and oblique shock angle. In wedge-stablized ODW, it largely determines the location where the chemical reactions are initiated behind the OSW, although the bounary layer effects are also not negligible [47]. With $\tau_i$, the induction length can be estimated from $L_i = u_s \cdot \tau_i$, corresponding to the reaction initiation location over a wedge [32]. Here $u_s$ is the gas velocity behind the OSW. When $L_i$ is less than the wedge length, post-OSW ignition can proceed. Typically, the ignition delay time $\tau_i$ can be obtained from 0D isochoric calculations, with the thermochemical states behind the OSW as initial conditions. These conditions are from the shock relations [32], with given flight altitude and Mach number. In this study, $\tau_i$ is defined as the duration from the initial to the instant when maximum heat release rate (HRR) happens in the ignition process [13].

Moreover, chemical excitation time, $\tau_e$, can quantify the rapidity of combustion heat release. Following Bradley et al. [48], $\tau_e$ can be defined as the time elapsed from the instant of 5% maximum HRR to that of the maximum from the autoignition process.

When we consider, for instance, the stoichiometric *n*-heptane / air mixture (droplet-free) with $H_0 = 30$ km and $M_0 = 10$, $\tau_i$ and $\tau_e$ from the 0D calculations are 21.7 and 2.3 µs, respectively, as seen from the HRR profile in Fig. 3(a). In previous studies on hydrogen ODW studies, different timescales are adopted to quantify the reaction progress. For example, in hydrocarbon fuel combustion, the heat release process is characterized by various key species, e.g., OH and CO [49]. Moreover, Figueria Da Silva et al. [33] define the total reaction time, $\tau_{i,1}$, as the duration to achieve 90% of the total temperature rise, $\Delta T$, during the autoignition process, while the heat release timescale, $\tau_{e,1}$, is the duration from the instant of 10%$\Delta T$ to that of 90%$\Delta T$. Accordingly, the chemical heat release time is the difference between them. With this appraoch, based on the temperature evolution in Fig. 3(b), the induction and heat release times are 29.4 and 9.9 µs, respectively, much longer than $\tau_i$ and $\tau_e$ annotated in Fig. 3(a). In an earlier study, Morris et al. [50] define the ignition delay $\tau_{i,2}$ as the total time reaching 50%$\Delta T$ and the heat release timescale $\tau_{e,2}$ is the time duration from 1%$\Delta T$ to 50%$\Delta T$. One can see from Fig. 3(b), they are resepctively 21.8 and 7.7 μs. If we compare the profiles of HRR and temperature in Fig. 3, we can see that the temperature-based criteria cannot reasonably define the chemical timescales for *n*-heptane autoignition, although they may be applicable for hydrogen autoignition [33,50]. We also examine other *n*-heptane/air mixtures with various flight altitudes or Mach numbers, and similar observations are observed.

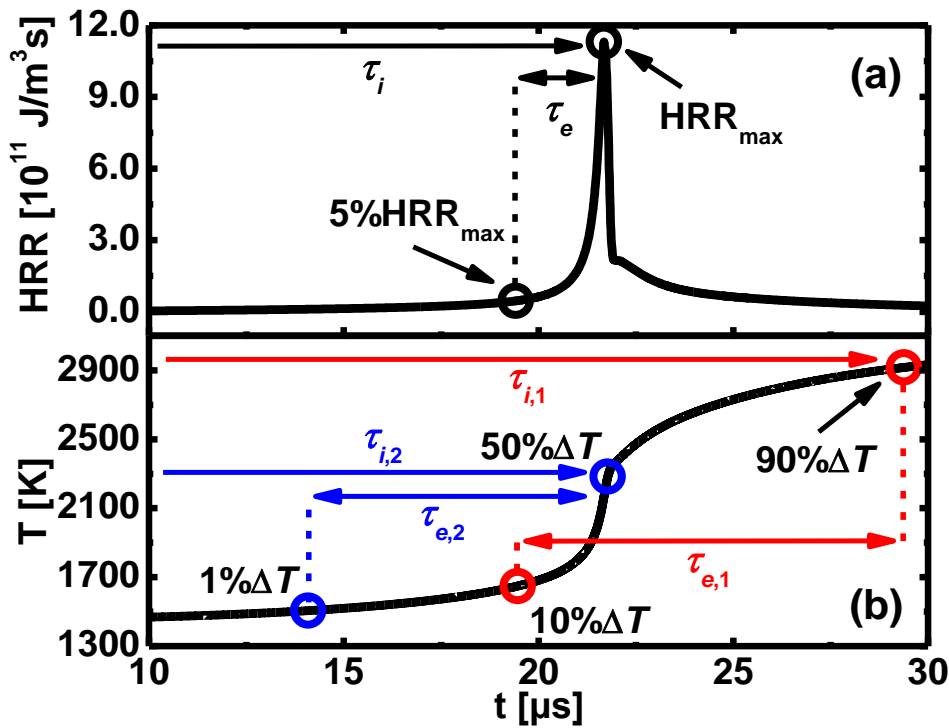

Figure 3: Time history of (a) heat release rate and (b) temperature in ignition of stoichiometric *n*-heptane/air gas mixtures. $H_0$ = 30 km and $M_0$ = 10.

## 4. Predictability of OSW-ODW transition mode

Whether the OSW-ODW transition mode can be accurately predicted is a signifcant topic in ODW studies. Figueria Da Silva et al. [33] attempt to identify a boundary between two modes based on the ratios of the induction to total reaction time, as defined in Section 3. They consider various pressures and temperatures of the shocked gas, and find that when $\tau_{e,1} \ll \tau_{i,1}$, the transition mode is abrupt, whilst when $\tau_{e,1} \approx \tau_{i,1}$, it is smooth. Moreover, Morris et al. [50] compare the OSW angle ($\beta_{\text{peak}}$) and equilibratium detonation wave angle ($\beta_{\text{eq}}$) and justify different transition modes based on the ratio of igniton time to equilibration time. When $\beta_{eq} \ll \beta_{peak}$, the transition mode is abrupt, which has an equilibration time much smaller than the corresponding ignition time, i.e., $\tau_{e,2} < 0.5(\tau_{i,2} - \tau_{e,2})$. When $\beta_{eq} \lesssim \beta_{peak}$, it is smooth, featured by the characteristic equilibration time satisfacying $0.5(\tau_{i,2} - \tau_{e,2}) \leq \tau_{e,2} \leq (\tau_{i,2} - \tau_{e,2})$. However, since their tempeature-based criterion may not accurately describe the ignition timescale, the accuracy of their method is questionable.

Recently, Teng et al. [51] propose a criterion to identify the transition mode based on the difference between OSW and ODW angles. They suggest that a smooth mode appears when the angle difference is less than 15° – 18°; otherwise, it is abrupt. The OSW/ODW angles are the superficial feature of oblique detonations. Miao et al. [14] investigate the predictability of different transition modes in ODWs. Their analysis shows that the transition modes can be characterized by the pressure ratio $P_d/P_s$, in which $P_d$ and $P_s$ are the pressures behind the ODW and OSW, respectively. When $P_d/P_s < 1.3$, a smooth transition appears; beyond that, an abrupt mode is observed. Moreover, a criterion based on the velocity ratio $\Phi = U_0/U_{CJ}$ ($U_0$ and $U_{CJ}$ are the the inflow velocity and the CJ speed, respectively) is also proposed by them to predict the transition modes [14]. For a given wedge angle $\theta$, the relation between the critical velocity ratio $\Phi^*$ and $\theta$ is also used for mode identification: when $\Phi < 0.98\Phi^*$, an abrupt transition occurs, whereas when $\Phi > 1.02\Phi^*$, it is smooth. Apparently, the above criteria are based on one or several flow parameters and therefore fail to correlate the ODW initiation with the nature of fuel chemistry under ODE-relevant conditions, which may become dominant when complicated fuels are considered.

The ignition delay time can well predict the post-OSW ignition location over the wedge surface, which have been coorborated by numerous studies (e.g., [10,16]). Moreover, the shorter chemical excitation time, the more rapid heat release. Fast energy depositon may significantly modulate local thermodynamic properties around the ignition spot and hence generate pressure or shock waves behind the OSW. These emanated waves of different intensities further compress the post-OSW gas and therefore lead to different ODW initiation modes [6,33]. This has been revealed by the chemical timescale from the chemical explosive mode analysis [22]. Figure 4 summarizes the possible influences of two timescales on gas ignition location and ODW initiation mode.

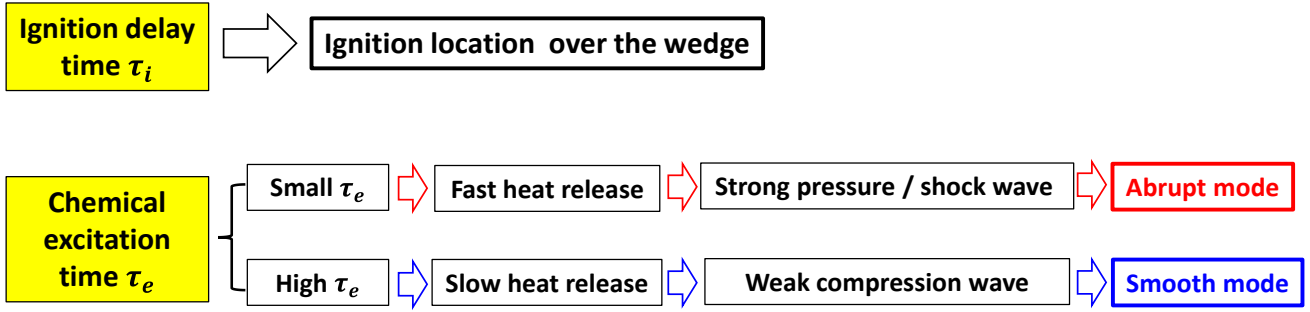

Figure 4: Influence of chemical timescales on ODW transition mode.

To corroborate the above analysis, Fig. 5 shows the chemical excitation time versus ignition delay time in numerical [10,13–16,33,49,52] and experimental [50,53] ODW studies with different fuels. Similar figure is also present in our work [13] but here we include more data to demonstrate the generality of the $\tau_e$ - $\tau_i$ relation. The transition mode is marked with open (abrupt) or solid (smooth) symbols. Their conditions are tabulated in Table 1. Note that the transition modes in Fig. 5 are recorded from the simulations or experiments in the references. Besides, the timescales are from 0D calculations (i.e., Eqs. 3 – 5) with the same chemical mechanisms as in the original work. The hydrogen mechanism by Burke et al. [54] is used for two experimental cases [50,53].

One can see that, for ODW's with hydrogen [10,14,33,49,52] or methane/hydrogen blends [15], smooth transition mostly occurs when $\tau_e = 0.2\tau_i \sim 0.5\tau_i$. Similar range for smooth transiton is also observed in the H$_2$ ODW experiments [50,53]. Furthermore, most acetylene ODW's with smooth transition [16] have around $\tau_e = 0.3\tau_i \sim 0.6\tau_i$. For *n*-heptane ODW's [13], smooth transition mode occurs roughly with $\tau_e = 0.5\tau_i \sim \tau_i$. In general, with smooth transition, the heat release is relatively slow in all the above fuels, i.e., the ratio of $\tau_e$ to $\tau_i$ is high. Also, based on Fig. 5, the abrupt mode (open symbols) is observed for smaller ratios of $\tau_e$ to $\tau_i$. Therefore, the applicability of ignition delay and chemical excitation timescales for the transition mode prediction can be confirmed. Moreover, another important implication from our analysis is that, to accurately predict the ODW behaviours, the chemical mechanism used should be able to correctly reproduce both ignition delay and chemical excitation time.

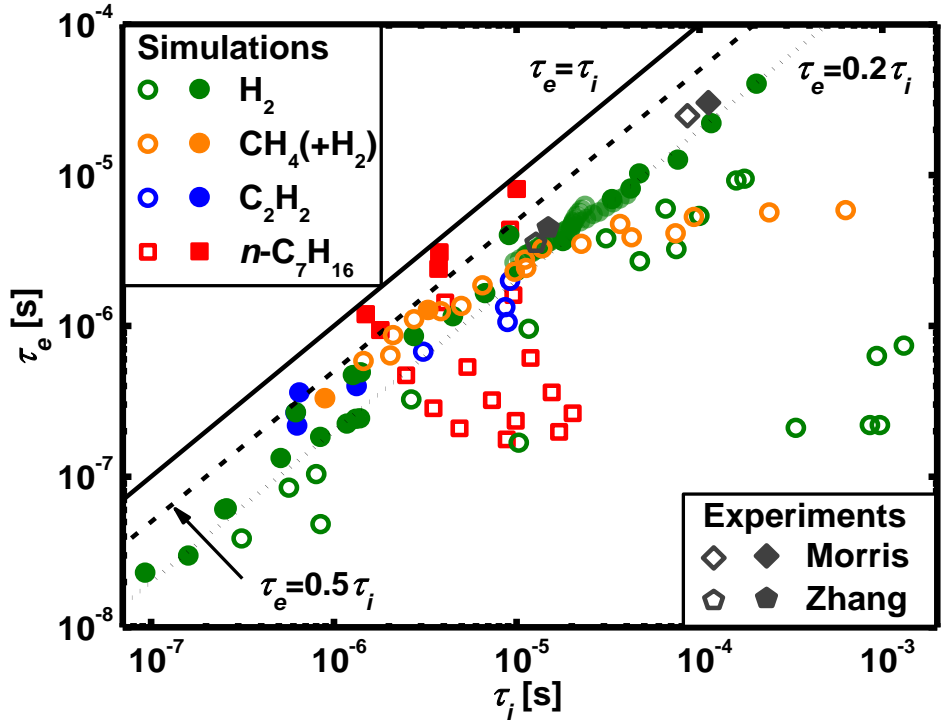

Figure 5: Chemical excitation time versus ignition delay time. Open/solid symbol: abrupt/smooth transition. Reference information for the data is in Table 1.

Table 1. Inflow conditions of oblique detonations in Fig. 5.

| Fuel | Equivalence ratio | Mach number | Pressure [atm] | Temperature [K] |
|---|---|---|---|---|
| $H_2$ [10,14,33,49,52] | 0.2 – 5.0 | 5.0 – 10.0 | 0.016 – 1.0 | 270 – 872 |
| $CH_4(+H_2)$ [15] | 0.0 – 1.0 (total=1.0) | 9.0 – 12.0 | 0.275 – 0.515 | 697 – 977 |
| $C_2H_2$ [16] | 1.0 | 7.0 – 10.0 | 0.2 | 298 |
| $n$-$C_7H_{16}$ [13] | 0.2 – 1.4 | 8.1 – 9.1 | 1.0 | 298 |
| Morris Exp. $H_2$ [50] | 1.0 | 5.85 | 0.118 | 282, 292 |
| Zhang Exp. $H_2$ [53] | 0.8, 1.0 | 6.6 | 0.0085, 0.0078 | 363, 348 |

## 5. Results and discussion

In this section, autoignition of $n$-heptane droplet/vapor/air two-phase mixtures will be studied. The ignition delay and chemical excitation time subject to various liquid fuel properties and flight conditions will be analyzed, to provide the insights into the initiation of liquid *n*-heptane ODW.

## 5.1 Autoignition of two-phase mixtures

Parametric studies are performed and four typical cases will be discussed first. Their details are tabulated in Table 2. The liquid ER is $\phi_l$ = 2.0, and the initial droplet diameter is $d_d^0$ = 10 μm. Two background gas ERs ($\phi_g$ = 0.4 and 1.0) and Mach numbers ($M_0$ = 9 and 10) are considered. Moreover, the initial reactor temperature and pressure range from 1,300 to 1,600 K and 3 to 10 atm, respectively, derived from the abovesaid freestream conditions. In the following analysis, we define the instant when droplet diameter $d_d$ is less than $10^{-8}$ m as the droplet evaporation time $\tau_{ev}$.

Table 2. Information of selected cases in Section 5.1.

| Case | $\phi_g$ | $\phi_l$ | $d_d^0$/μm | $H_0$/km | $M_0$ | Initial reactor temperature/T | Initial reactor pressure/Pa |
|---|---|---|---|---|---|---|---|
| 1 | 0.4 | 2.0 | 10 | 21 | 9 | 1339.9 | 947593.3 |
| 2 | 1.0 | | | | | 1261.3 | 977590.7 |
| 3 | 0.4 | | | 30 | 10 | 1589.6 | 322768.4 |
| 4 | 1.0 | | | | | 1490.1 | 334198.9 |

Figure 6 shows the change of the droplet temperature/diameter, gas temperature, HRR, and species mass fraction in cases 1 and 2. In these cases, the droplets are fully vaporized at $\tau_{ev}$ = 0.099 ms with $\phi_g$ = 0.4, and 0.107 ms with $\phi_g$ = 1.0; see the curves of $d_d$~$t$ in Fig. 6. Both are one order of magnitude smaller than the respective ignition delay time. Therefore, although the initial mixtures are heterogeneous with *n*-heptane droplets and vapor, nonetheless, majority of the ignition process, including thermal runaway and part of the radical runaway, actually proceeds in a homogeneous (gaseous) mixture.

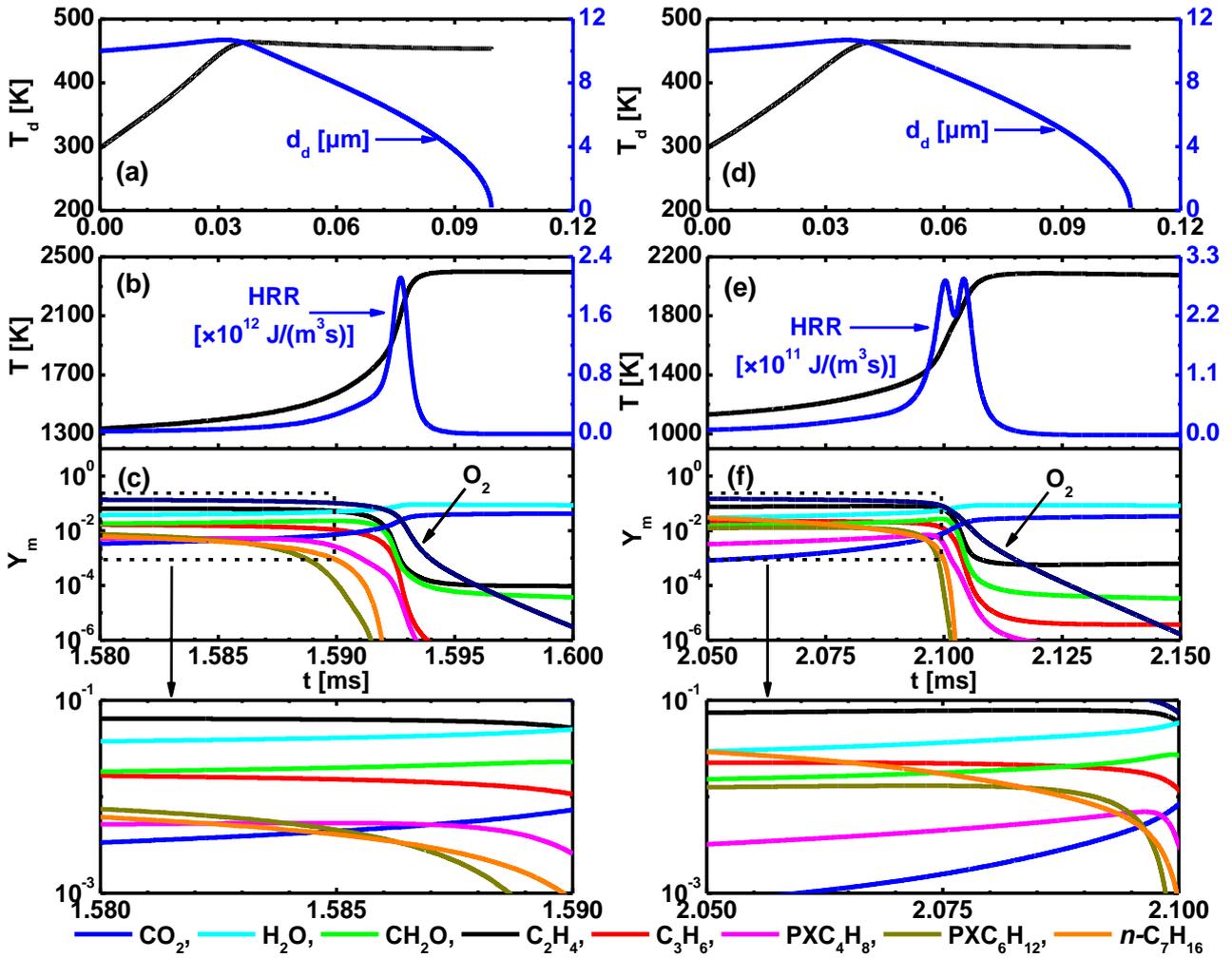

Figure 6: Time evolutions of droplet temperature, diameter, gas temperature, HRR, and species mass fractions with $\phi_g = 0.4$ (left column, case 1) and $\phi_g = 1.0$ (right column, case 2).

In both cases, as the chemical reactions proceed, the gas temperature gradually increases during the chemical runaway process, as indicated in Figs. 6(b) and 6(e). When the thermal runaway occurs, the rapid temperature rise can be found and then the corresponding equilibrium states are reached. For instance, the equilibrium temperature is 2355.5 K and 2015.5 K for the $\phi_g = 0.4$ and 1.0 cases, respectively. Strong heat release starts respectively at 1.588 ms and 2.067 ms in two cases, and their corresponding ignition delay is $\tau_i = 1.593$ ms and 2.104 ms, respectively. The former $\tau_i$ is shorter than the latter, which is reasonable since full vaproization of the liquid fuel leads to a richer composition in case 2. However, the chemical excitation time is 5.18 μs and 37.36 μs for cases 1 and 2, due to the endothermic and pyrolysis of excess fuel.

Change of the species mass fractions in Figs. 6(c) and 6(f) is mainly affected by gas chemistry, since droplets are fully gasified well before ignition occurs in both cases. In the induction stage, the $n$-$C_7H_{16}$ mass fraction decreases slowly, accompanied by increased concentration of small hydrocarbon species, such as $CH_2O$, $C_2H_4$ and $C_3H_6$. This can be more clearly seen from the zoom-in panels in the last row of Fig. 6. They react with the oxygen, and hence the $O_2$ mass fraction decreases. The corresponding HRR reaches the peak when the thermal runaway occurs, which correspponds to the ignition delay time. After the thermal runaway, the concentrations of most intermediate species are significantly reduced. Note that two peaks of the HRR in case 2 are caused by a short-period chemical runaway during the heat release, related to the $C_2H_4$ chemistry, which will be clearly revealed in Fig. 11(e) through analyzing the chemical explosive mode.

In the following, we will further study the scenarios in which the droplet evaporation directly interacts with the gas autoignition, i.e., $\tau_{ev}$ is comparable to or longer than $\tau_i$. They are cases 3 and 4 in Table 2, with the gas ERs of $\phi_g$ = 0.4 and 1.0, respectively. The rest parameters are ($\phi_l$, $d_d^0$, $H_0$, $M_0$) = (2.0, 10 μm, 30 km, 10). Figure 7 shows the time evolutions of the same variables as in Fig. 6. For case 3, the droplets complete the evaporation at $\tau_{ev} \approx 0.052$ ms; see the $d_d$~$t$ curve in Fig. 7(a). It is longer than the ignition delay time, $\tau_i = 0.0071$ ms, as seen from the HRR curve in Fig 7(b). During the autoignition process, the droplets in the reactor are heated by surrounding high-temperature gas, which concurrently increases from its initial 1589.6 K due to combustion heat release. This results in increased droplet temperature and diameter before 0.012 ms, see Fig. 7(a). Note that slight droplet expansion (by 6.4%) is because the liquid $n$-heptane density decreases with the droplet temperature. However, during this period, droplet evaporation is still weak, comfirmed by the limited amount of fuel vapor release (not shown in Fig. 7). During the droplet heating/expansion stage, thermal runaway of the gas chemistry occurs at 0.0071 ms, in a two-phase mixture. In this senario, the ignition process, including the chemical and thermal runaway, is mainly influenced by droplet heating (thermal effects), and the fuel vapor addition plays a minor role.

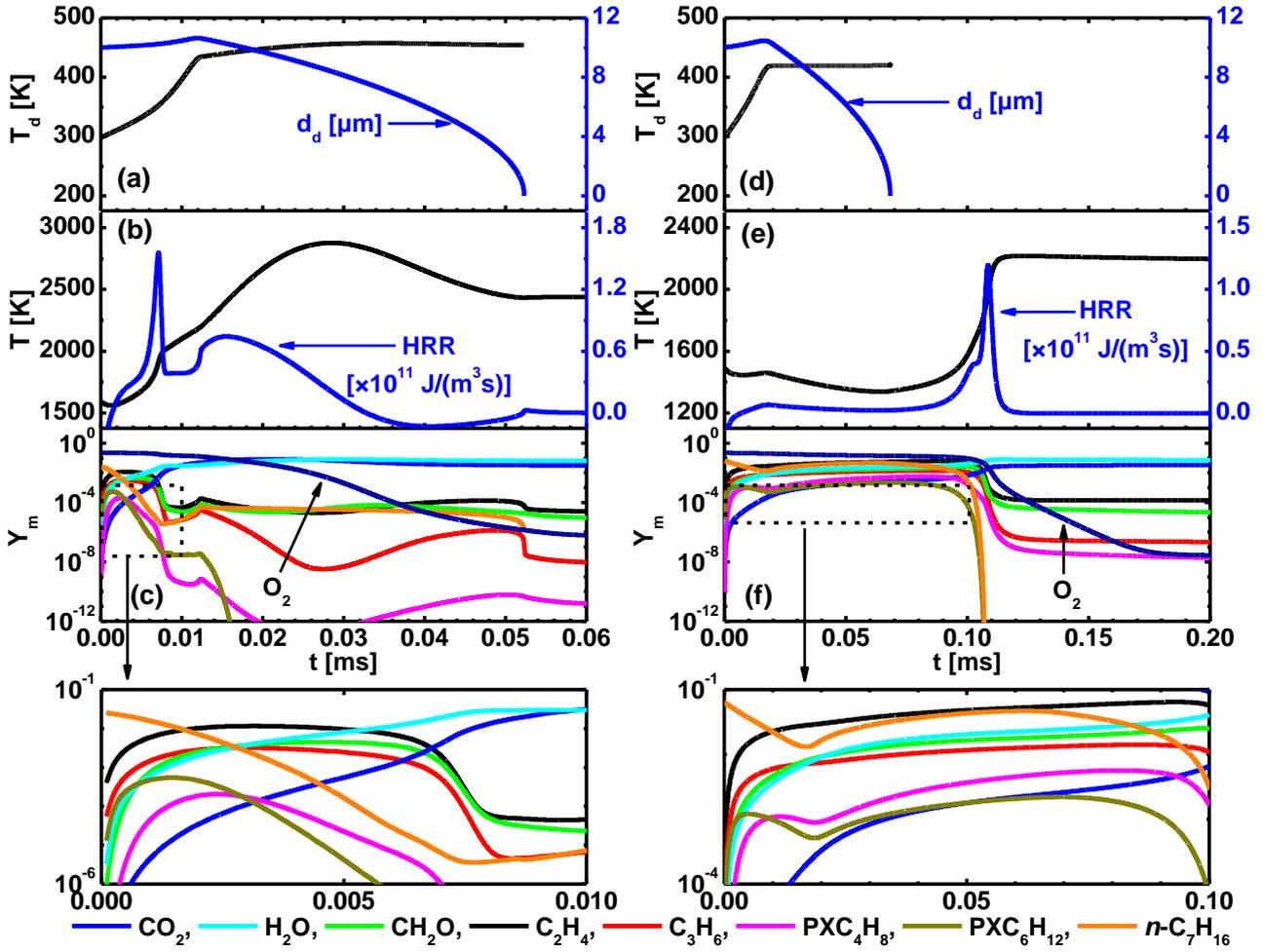

Figure 7: Time evolutions of droplet temperature, diameter, gas temperature, HRR, and species mass fractions with $\phi_g = 0.4$ (left column, case 3) and $\phi_g = 1.0$ (right column, case 4).

At the post-ignition stage, the droplets continue to absorb heat for phase change from the high-temperature background gas. Meanwhile, the combustion HRR decreases rapidly due to depletion of the fuel. At 0.012 ms, the droplet diameter begins to decrease, which indicates that strong droplet evaporation is initiated, and $n$-heptane vapor is hence released into the gas phase. Due to the residual oxygen in the reactor, secondary ignition occurs, accompanying the on-going $n$-heptane droplet evaporation. With this, the combustion HRR rises again, leading to a continuously increased temperature towards 0.03 ms, see Fig. 7(b). As the oxygen is gradually consumed, the HRR decreases, and the oxygen mass fraction is reduced to below $5 \times 10^{-4}$ at 0.03 ms. Meanwhile, the released $n$-heptane vapor first decreases and then increases in the reactor, which would be continuously pyrolyzed in the post-ignition atmohsphere. This leads to increased mass fractions of intermediate species, e.g., $C_3H_6$ and $PXC_4H_8$, see Fig. 7(c). The gas temperature gradually decreases beyond 0.03 ms, beucase of the enothermonic nature of fuel pyrolysis. After induction process, the intermediate species are gradually consumed. Since 0.052 ms, the temperature and mass fraction, e.g., $O_2$, $CO_2$ and $H_2O$, gradually relax towards equilibrium state in the reactor.

For the stoichiometric case 4 in the right column of Fig. 7, the ignition delay time is $\tau_i = 0.108$ ms, much longer than that of case 3. Nonetheless, the droplets completely vaproize at $\tau_{ev} = 0.068$ ms, in the midst of the chemical runaway process. This implies that the droplets may have both thermal (i.e., energy absorption) and kinetic (i.e., fuel vapor addition) effects on the evolution of gas chemistry. In fact, one can see from Figs. 7(e) and 7(f) that during the evaporation stage, the gas temperature decreases due to the energy absorption by the droplets, and at 0.068 ms the gas temperature is lowest (1337.9 K, reduce by 10.2% relative to the initial 1490.1 K) before ignition happens. As the fuel vapor is added, the fuel decomposes and the mass fractions of intermediate species, e.g., $C_2H_4$, $C_3H_6$ and $PXC_4H_8$, increase. Both droplet evaporation and *n*-heptane pyrolysis absorb heat from the gas phase and hence jointly reduce the gas temperature. After 0.068 ms, the system gradually proceeds towards thermal runaway, featured by quickly increased HRR and gas temperature, as well as reacted fuel and intermediate species, see Figs. 7(e) and 7(f).

**5.2 Heat and mass transfer**

Figure 8 demonstrates the heat and mass transfer ($\Delta E$ and $\Delta m$) between the evaporating droplets and igniting two-phase mixtures in cases 1-4. In Fig. 8(a), before 0.03 ms, $|\Delta E|$ gradually increases, but $\Delta m$ is small due to limited evaporation. After that, the evaporation starts to increase rapidly ($\Delta m \uparrow$) and $|\Delta E|$ decreases quickly. It features the beginning of the rapid evaporation process. At the end stage of the rapid evaporation ($t > 0.037$ ms), $|\Delta E|$ changes from negative to positive values. According to Eq. (9), the rapid evaporation of droplet results in a higher enthalpy change rate, $-\dot{m}_{d,i} h_g$, which can also be confirmed from the mass transfer curve in Fig. 8(a). The mass transfer slows down and $\Delta m$ reaches around zero at around 0.099 ms. Also, $\Delta E$ gradually decreases to zero, which indicates that the droplet evaporates completely. The results in cases 2 (i.e., Fig. 8b) are qualitatively similar to those in Fig. 8(a).

In Fig. 8(c), the mass transfer is limited throughout the whole autoignition period (annotated with $\tau_i$ in Fig. 8c). Rapid evaporation begins after the mixture is ignited, with quickly increased $\Delta m$ and $|\Delta E|$. The vaproized *n*-heptane is pyrolysed with considerable heat absorption, and accordingly the interphase heat transfer is weakened ($|\Delta E| \downarrow$) between 0.0106 and 0.0137 ms. Subsequently, combined with Fig. 7(b), it can be seen that addition of the fuel vapor results in continuous combustion and temperature rise, and $|\Delta E|$ increases before 0.025 ms. After that, $|\Delta E|$ is gradually reduced, leading to limited evaporation ($\Delta m \downarrow$). For Fig. 8(d), droplet evaporation occurs almost concurrently with the autoignition process. At the droplet heating stage ($t < 0.015$ ms), $|\Delta E| \uparrow$ with small $\Delta m$. Rapid evaporation follows ($|\Delta E| \downarrow$, $\Delta m \uparrow$) until the droplets are fully gasified at around 0.07 ms ($|\Delta E| \to 0$, $\Delta m \to 0$).

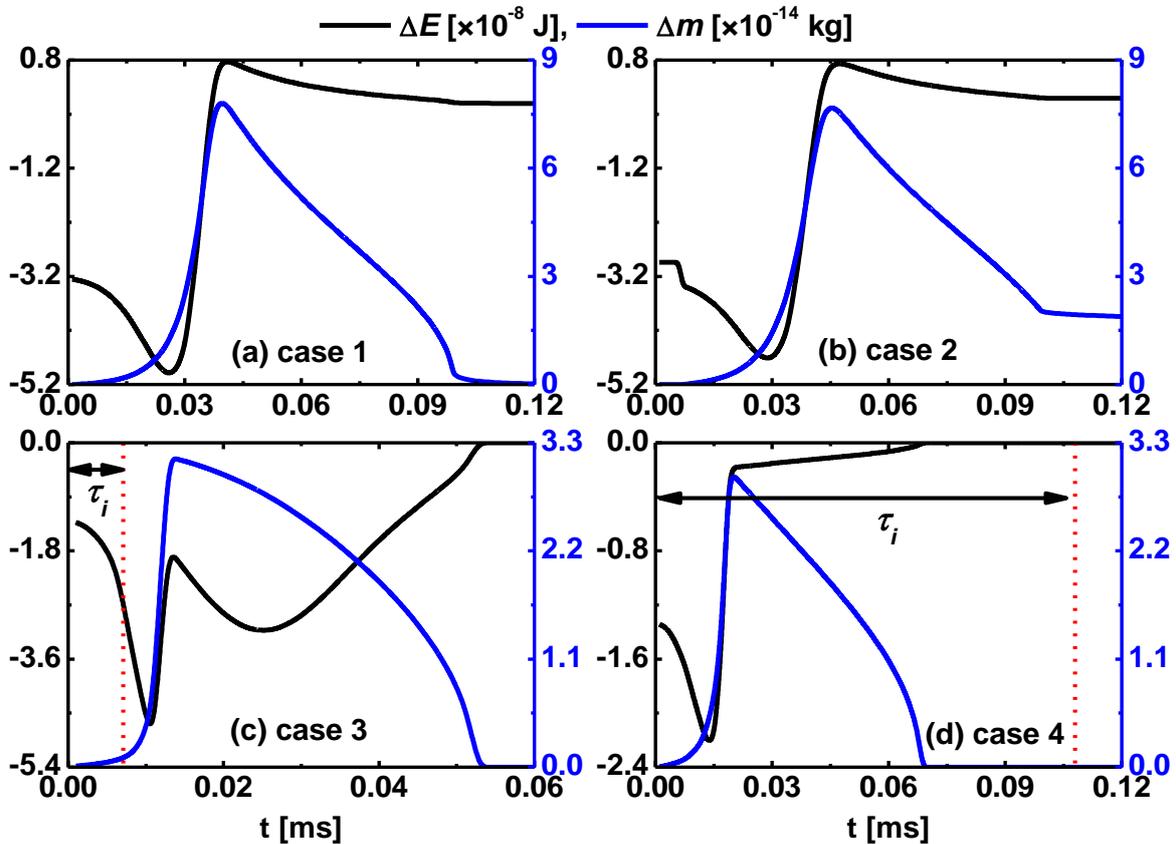

Figure 8: Time evolutions of interphase heat and mass transfer in cases 1-4.

Comparing cases 1-4, we can see that the instant of the beginning of rapid evaporation in stoichiometric ratio is slightly delayed than in $\phi_g = 0.4$, and the total evaporation time of droplet is longer. This is related to the decrease in reactor initial temperature and the increase in pressure.

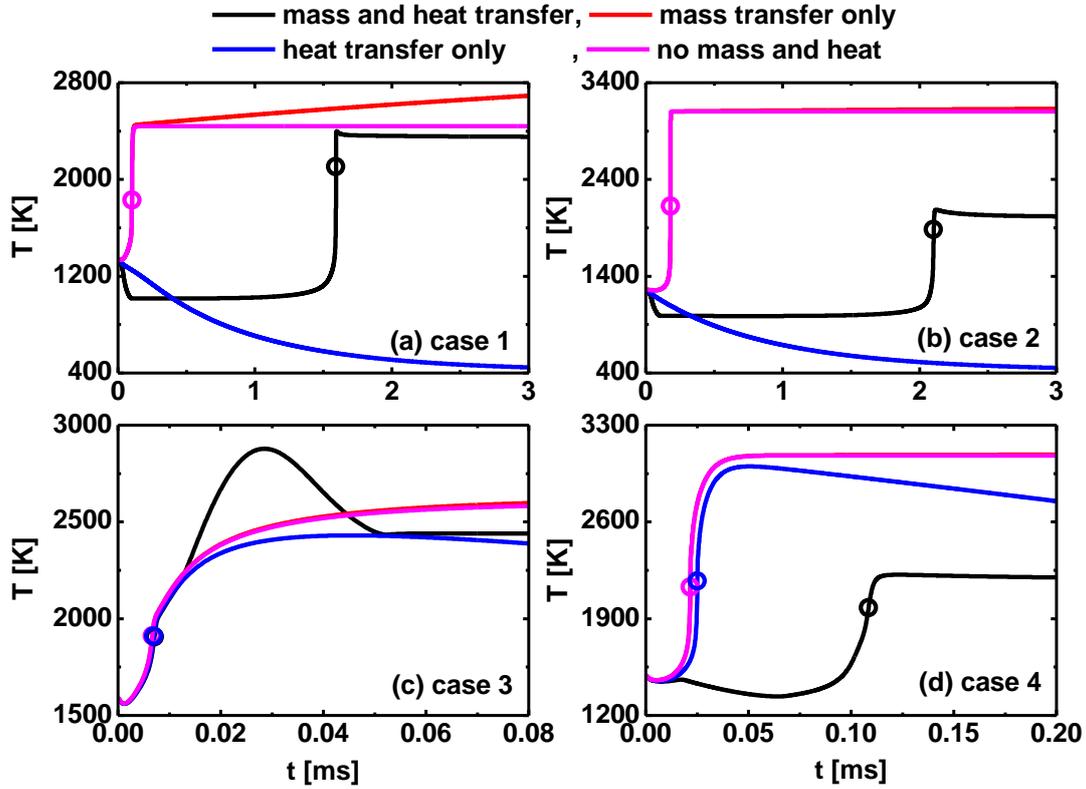

Figure 9: Time evolutions of gas temperature with or without mass/heat transfer in cases 1-4. Circle: ignition delay time.

To elucidate how the heat and mass transfer influences the autoignition of two-phase mixtures, we conduct numerical experiments by turning off droplet evaporation and/or heat exchange models. Figure 9 shows the time evolutions of gas temperature in cases 1-4, and the corresponding ignition delay time comparisons are given in Fig. 10. We first look at case 1. It can be seen from Fig. 9(a) that, without considering the heat transfer (red line), the interphase mass transfer has limited influences on the autoignition transient, and $\tau_i$ is close to that of the gas-only mixture (both mass and heat transfer are off, see Fig. 10). Nonetheless, the heat transfer from the evaporating droplets significantly delays the autoignition of $n$-heptane vapor. Specifically, if only heat transfer is retained, autoignition fails ($\tau_i \to +\infty$), as shown in Figs. 9(a) and 10. When both heat and mass transfer are considered, successful ignition can proceed, which confirms the promotive role of fuel vapour addition in a lean background gas. Similar tendency is found in case 2 in Figs. 9(b) and 10. In cases 3-4 with direct interaction between droplet evaporation and gas autoignition, if only heat transfer is retained, this would not lead to autoignition failure. When the ignition delay time is significantly smaller than the droplet evaporation time (case 3 in Fig. 9c), the droplet mass and heat transfer have little effect on $\tau_i$, see Fig 10. Otherwise, the autoignition of *n*-heptane vapor is significantly delayed (e.g., case 4 in Fig. 9d).

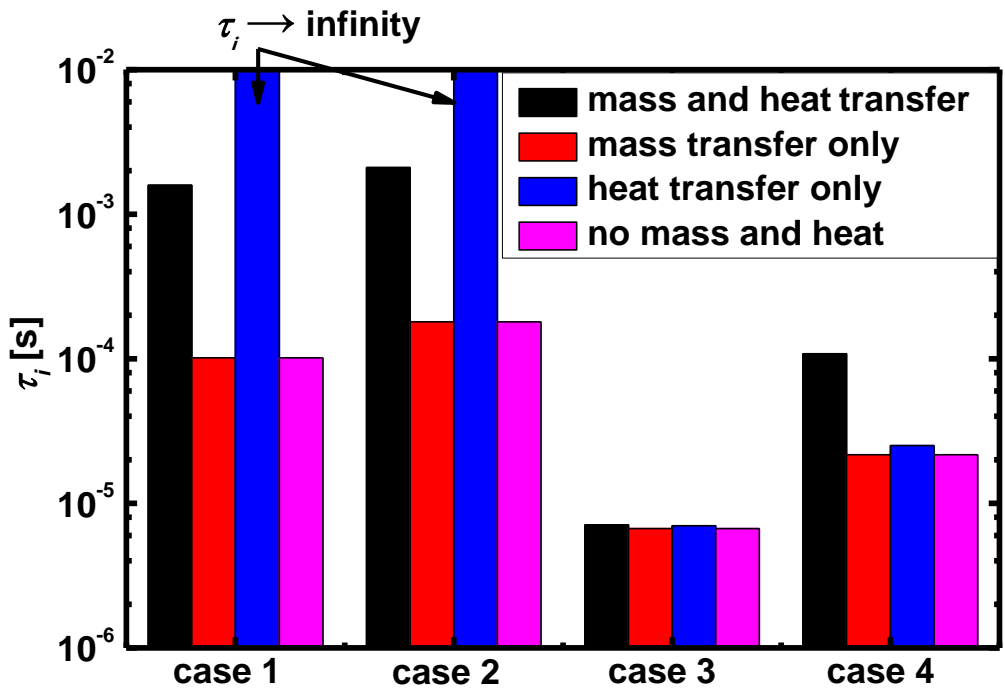

Figure 10: Ignition delay time with different interphase coupling mechanisms in cases 1-4.

### 5.3 Chemical explosive mode analysis

The chemical explosive mode analysis [55–57] is employed to analyze the underlying chemistry in autoignition of two-phase mixture in cases 1-4. Details of this method are available in Refs. [56–58]. Figures 11 and 12 show the time history of chemical explosive mode ($\lambda_{\text{CEM}}$), explosion index for temperature ($EI\_T$), explosion indices for species ($EI\_Y_m$), and reaction participation indices (PI). The definitions of the explosion and participation indices can be found in Ref. [56]. To re-iterate, $\lambda_{CEM} > 0$ means that the mixture tends to explode, whilst $\lambda_{CEM} < 0$ means that it is burnt or fails to ignite. The condition of $\lambda_{CEM} = 0$ therefore separates the CEM and post-combustion or inert mixing regions [56]. Moreover, the contributions of temperature or species to the CEM can be evaluated with the Explosion Index (EI). The local chemical reaction is dominated by chain-branching reactions if the radical EI is high, whilst dominated by thermal runaway if the temperature plays an important role [56]. The PI quantifies the contribution of individual reactions to the CEM.

We first analyze case 1, in Figs. 11(a) – 11(c), which correspond to Figs. 6(a)-6(c). Before ignition (i.e., 1.593 ms), the gas temperature is dominant in the chemical explosive mode (featured by high $EI\_T$ in Fig. 11a), but the role of the intermediate species (e.g., $CH_2O$, $C_2H_4$ and $C_3H_6$) is also gradually increased towards the thermal runaway at 1.593 ms. R100: $NXC_7H_{16}+OH \leftrightarrow SXC_7H_{15}+H_2O$ and R85: $PXC_4H_9 \leftrightarrow C_2H_5+C_2H_4$ contribute more for $t < 1.585$ ms. Subsequently, the reaction R1: $H+O_2 \leftrightarrow OH+O$ becomes dominant, providing the significant amount of OH radical. In this case, R1: $H+O_2 \leftrightarrow OH+O$ and R8: $H+O_2+M \leftrightarrow HO_2+M$ contribute more to the heat release process, see Fig. 11(c). When $t > 1.593$ ms, the role of the CO species increases, and R17: $CO+OH \leftrightarrow CO_2+H$ contributes to the conversion of CO to $CO_2$.

For case 2 in Figs. 11(d) – 11(f), $\lambda_{CEM}$ becomes negative at 2.105 ms, consistent with the ignition delay time from Fig. 6. Similar to the results in Fig. 11(a), the temperature is relatively important for the chemical explsove mode before ignition. One can see from Fig. 11(f) that R100: $NXC_7H_{16}+OH \leftrightarrow SXC_7H_{15}+H_2O$ and R85: $PXC_4H_9 \leftrightarrow C_2H_5+C_2H_4$ contribute more before 2.102 ms. Between the two heat release peaks (see Fig. 6e), the $EI\_T$ quickly decreases, instead, the EI of $C_2H_4$ become high, indicating an instenenous radical runaway period. Besides, the small hydrocarbon cpncentrations, like $CH_2O$ and $C_3H_6$, also increase, see Fig. 11(e). Therefore, large hydrocarbon species dominate in the early stage of the radical runaway process (before 2.102 ms), whilst smaller species become more significant at the later stage. Besides, when the HRR reaches the peak around 2.105 ms, the reaction R1: $H+O_2 \leftrightarrow OH+O$ becomes dominant (see Fig. 11f), providing the significant amount of OH radical to oxidize the CO to $CO_2$ in R17: CO+OH ↔ $CO_2$+H.

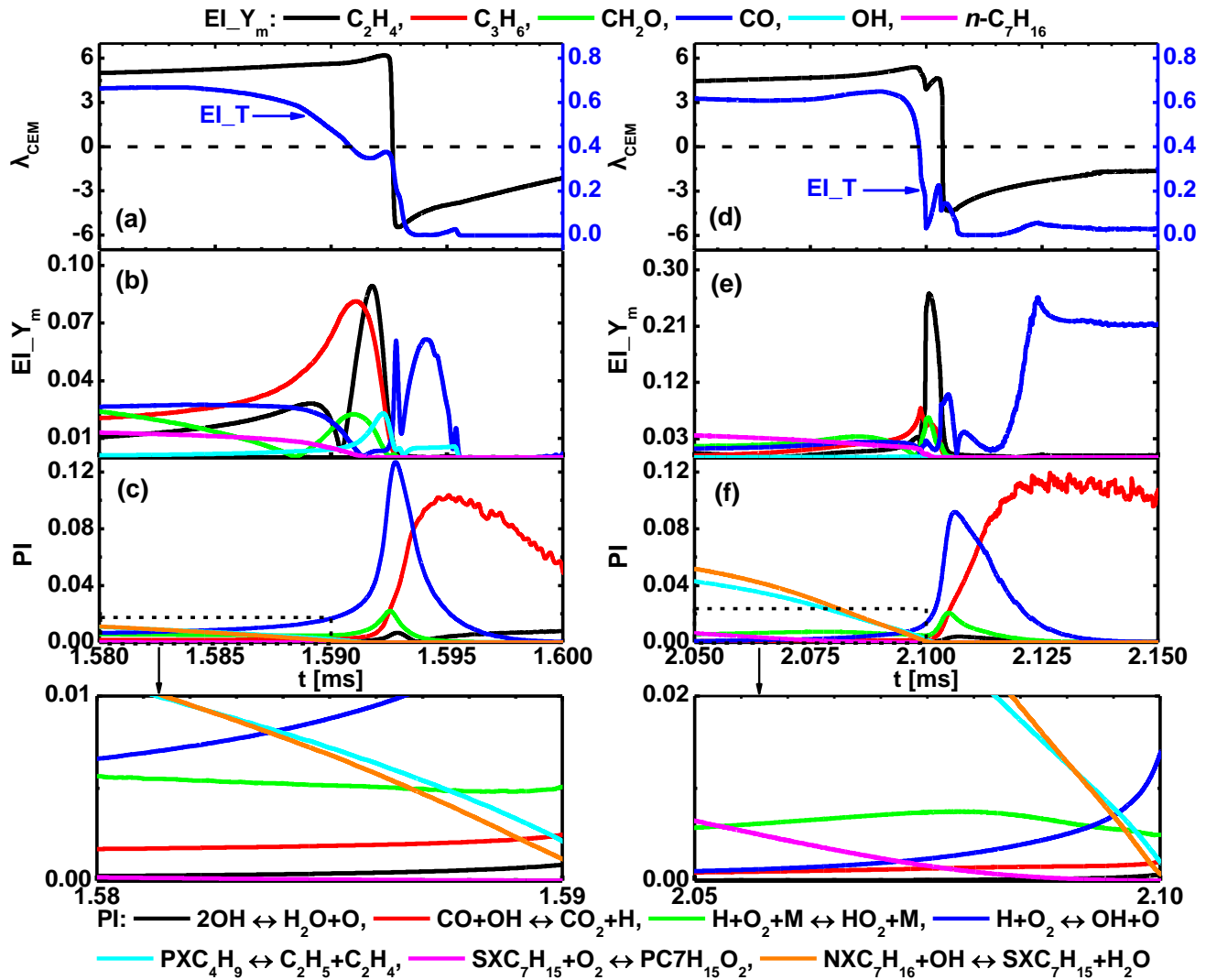

Figure 11: Time history of chemical explosive mode, explosion indices for temperature / species, and participation indices with $\phi_g = 0.4$ (left column, case 1) and $\phi_g = 1.0$ (right column, case 2). Dashed line: $\lambda_{CEM} = 0$.

Figures 12(a) and 12(d) respectively show the CEM results in cases 3 and 4 in Figs. 7(a) and 7(d). It can be seen from Fig. 12(a) that the intrinsic explosion propensity of the gaseous mixture changes alternately with time, due to the exisntence of evaporating droplets in the reactor, which is not observed from other cases. This corresponds to multiple zero-crossing points of the $\lambda_{CEM}$ curve, separating the explosive and non-explosive states [56]. Specifically, there are three durations with $\lambda_{CEM} > 0$: *A*: $t = 0 – 0.0076$ ms, *B*: $t = 0.0295 – 0{,}0381$ ms and *C*: $t > 0.0523$ ms. Note that the latter two are in the post-ignition period. For duration *A*, the mixtures have the strongest explosive properties (highest $\lambda_{CEM}$), and the sensitivity of the chemical reactions changes from temperature-controlled state to OH-controlled one, as revealed from Figs. 12(a) and 12(b). Meanwhile, R100: $NXC_7H_{16}+OH \leftrightarrow SXC_7H_{15}+H_2O$, R85: $PXC_4H_9 \leftrightarrow C_2H_5+C_2H_4$, and R1: $H+O_2 \leftrightarrow OH+O$ contribute more to the CEM during this period. R100 and R85 are respectively related to the initial oxidation of fuel species and decomposition of intermediate hydroaron species, whilst R1 indicates the generation of key radicals of OH and O from chain-branching reactions.

In duration *B*, since *n*-heptane droplets keep vaporizing, the gas mixtures restore to a highly explosive state with sufficient oxygen (see Fig. 7c). In this duration, Fig. 12(b) shows that OH and CO species have more contributions. For duration *C*, the $\lambda_{CEM}$ is 2-3 orders of magnitude lower compared to those of the preceding two durations, as observed in Fig. 12(a). This is reasonable, because of depletion of evaporating droplets, fuel vapor, and also oxygen. Apparnetly, the highest EI is from CO species, whilst the highest PI is from R17: $CO+OH \leftrightarrow CO_2+H$.

For case 4, the droplets complete the evaporation at around 0.07 ms, ahead of the autoignition delay time. For most of the time before autoignition, the gas temperature EI is high. This indicates the nature of the lengthened thermal runaway process in this event. Meanwhile, the EI of *n*-$C_7H_{16}$ maintains at a high level, and after the droplets are fully vaproized, its value is gradually reduced. This signifies the the influences of the released fuel vapor on the gas chemistry. Moreover, R100: $NXC_7H_{16}+OH \leftrightarrow SXC_7H_{15}+H_2O$ and R85: $PXC_4H_9 \leftrightarrow C_2H_5+C_2H_4$ play important roles during this period, and this is consistent with the importance of the elementary reactions prior to ignition in Figs. 12(a)-12(c). It should be noted that the $\lambda_{CEM}$ decreases, albeit slightly, as the droplet evaporation proceeds, and gradually increases after complete droplet vaporization (around 0.07 ms, see Fig. 12d), which may be because of the absence of the latent of heat asborption from the gas phase and evolvtion of the underlying evolutions of gas chemistry.

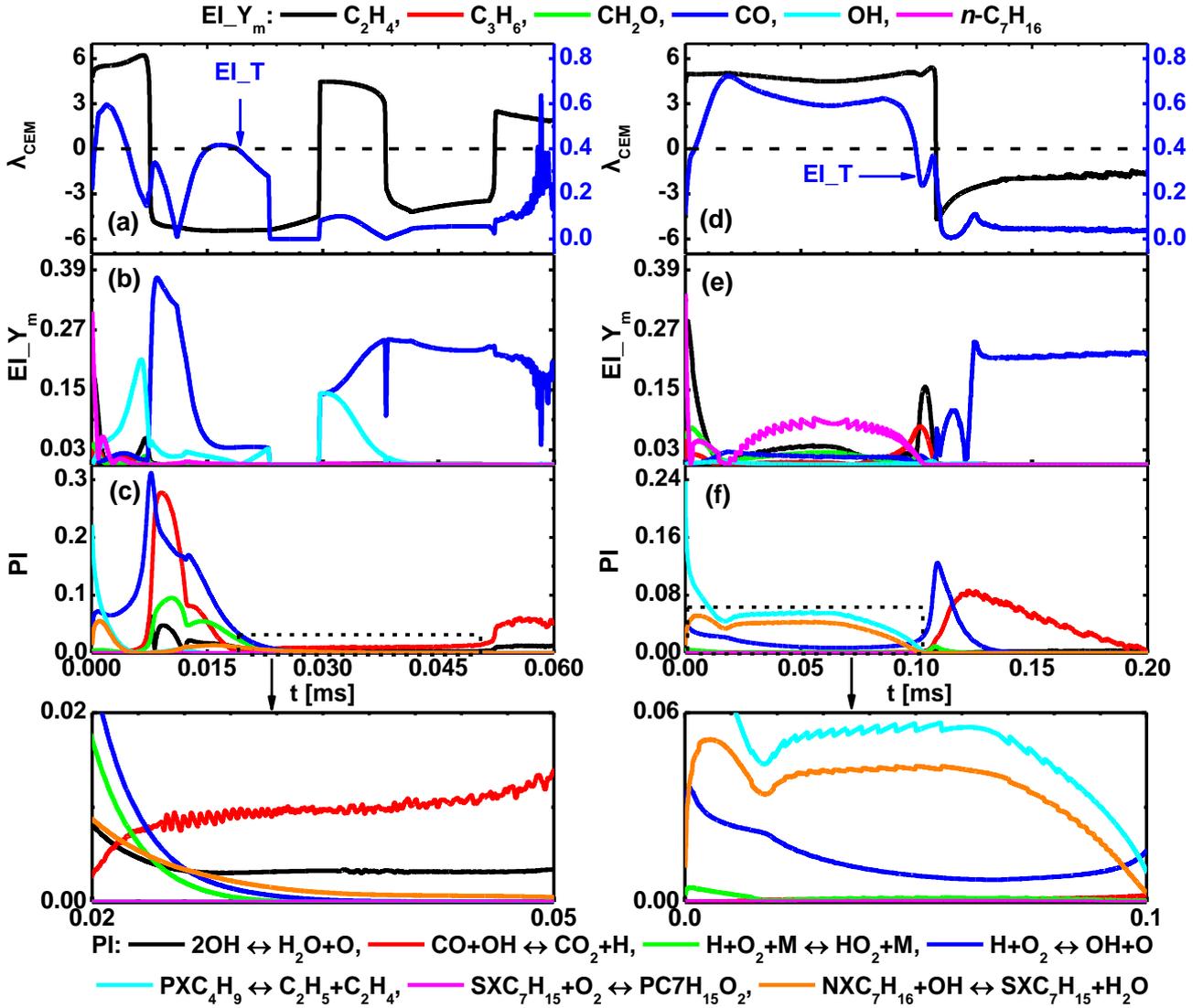

Figure 12: Time history of chemical explosive mode, explosion indices for temperature / species, and participation indices with $\phi_g = 0.4$ (left column, case 3) and $\phi_g = 1.0$ (right column, case 4). Dashed line: $\lambda_{\text{CEM}} = 0$.

### 5.4 Homogeneous versus heterogeneous ignition: flight condition effects

The mixture compostion (i.e., two-phase or gas-only) when the autoignition happens would directly influence the combustion heat release and therefore it is of great importance to understand how the flight conditions affect it. The timescales of autoignition and droplet evaporation under various flight altitudes $H_0$ and Mach number $M_0$ are presented in Fig. 13. In this section, the liquid $n$-heptane ER is fixed to be $\phi_l = 1.0$ and the initial droplet diameter is $d_d^0 = 10$ μm. Two background gas ERs are considered, i.e., $\phi_g = 0.4$ (Figs. 13a-13b) and 1.0 (Figs. 13c-13d). $M_0 = 9$ is fixed for Figs. 13(a) and 13(c), whilst $H_0 = 25$ km for Figs. 13(b) and 13(d).

In Figs. 13(a) and 13(c), the ignition delay time is generally greater than the evaporation time at respective altitudes (i.e., $\tau_i > \tau_{ev}$), and hence majority of the autoignition proceeds in a homogeneous (gaseous) mixture. Moreover, the ignition delay time increases first and then decreases with the flight altitude $H_0$. This is because the inflow temperature increases approximately linearly but the pressure decreases exponentially as the altitude $H_0$ increases, see Fig. 1(b), which leads to the change of the initial reactor conditions accordingly. The ignition delay time reaches the maximum when $H_0$ is 31 km when the background gas ER is $\phi_g = 0.4$. For $\phi_g = 1.0$ in Fig. 13(c), the longest ignition occurs at $H_0 = 33$ km. Also, the droplet evaporation time is shorter with the altitude in both Figs. 13(a) and 13(c), which is because the gas temperature increases as $H_0$ increases when $M_0$ is fixed.

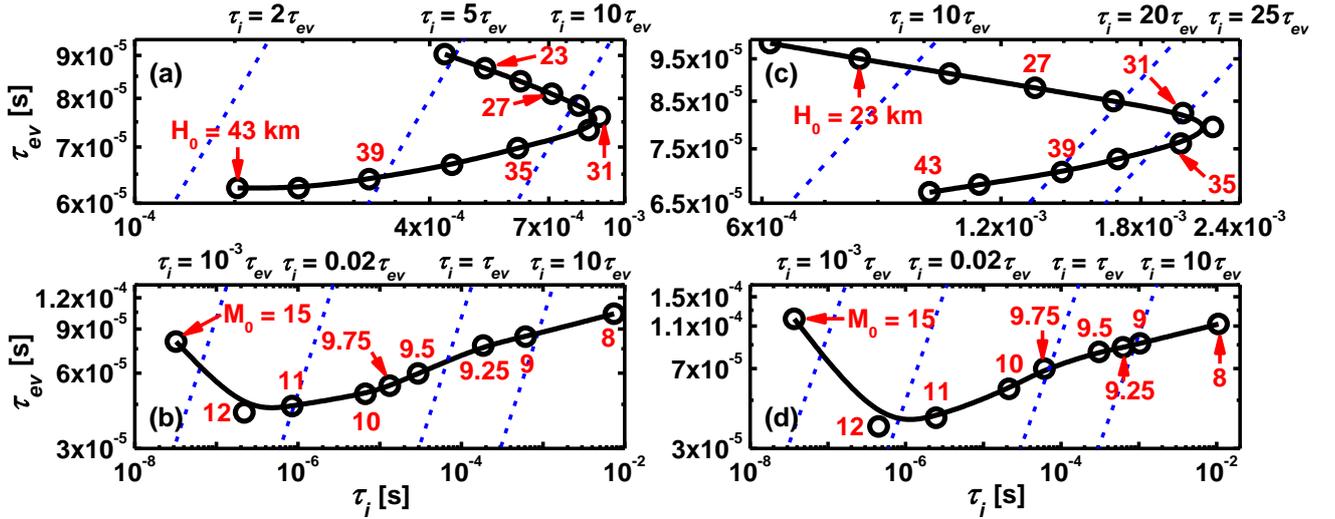

Figure 13: Ignition delay time versus droplet evaporation time with different flight altitudes and Mach numbers: (a-b) $\phi_g = 0.4$ and (c-d) $\phi_g = 1.0$. Dashed lines indicate different relations between $\tau_i$ and $\tau_{ev}$. $M_0 = 9$ for (a) and (c), $H_0 = 25$ km for (b) and (d). $\phi_l = 1.0$ and $d_d^0 = 10$ μm.

One can see from Fig. 13(b) and 13(d) that, when $H_0 = 25$ km, the ignition delay time is longer than the droplet evaporation time when the flight Mach number $M_0$ is lower (i.e., $\tau_i > \tau_{ev}$, when $M_0 < 9.5$ for $\phi_g = 0.4$ and $M_0 < 9.75$ for $\phi_g = 1.0$). They correspond to the homogeneous ignition. In these cases, the OSW intensity is relatively weak. Besides, the thermodynamic state (e.g, $T$ and $p$) of the gaseous mixtures behind OSW are small. Thus, the initial reactor pressure or temperature are lower. The ignition delay time decreases quickly with increased $M_0$. Similar tendency is observed for $\tau_{ev}$. However, for $M_0 < 12$, $\tau_{ev}$ increases with $M_0$. When $M_0$ is 15, the evaporation time is prolonged significantly. This is because high surrounding pressure induced by strong OSW inhibits droplet gasification. If one furthre compares the results from the left and right column in Fig. 13, the ignition delay time, evaporation time, and range of homogeneous ignition generally increase with the background gas ER $\phi_g$, which is mainly caused by the lower initial reactor temperature when $\phi_g = 1.0$.

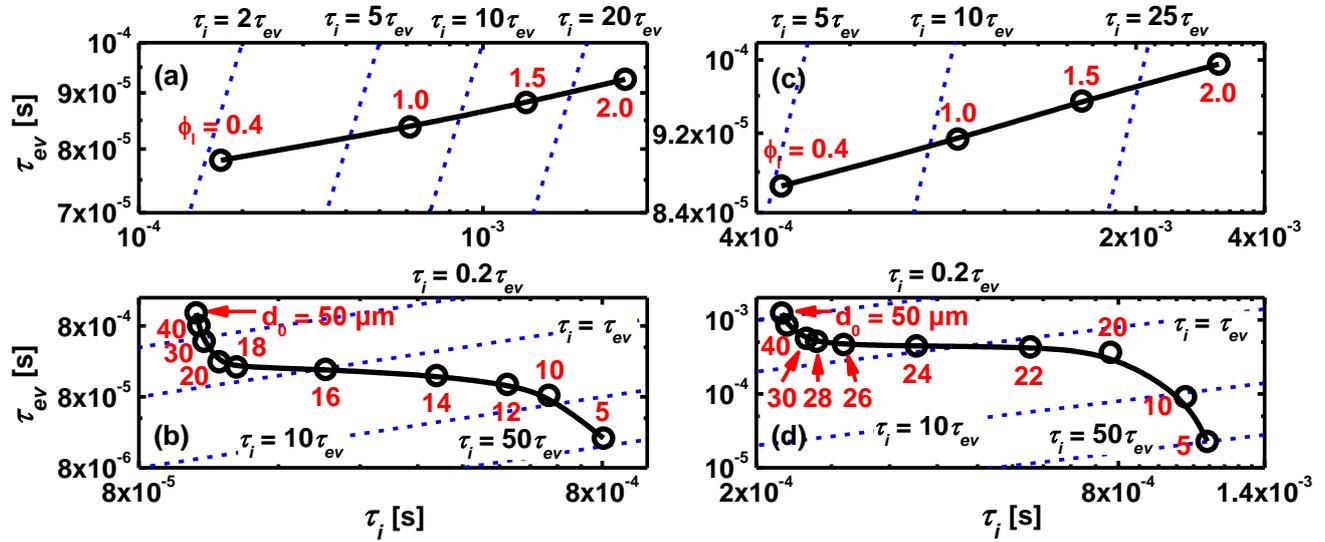

Figure 14: Ignition delay time versus droplet evaporation time with different liquid ERs and droplet diameters: (a-b) $\phi_g = 0.4$ and (c-d) $\phi_g = 1.0$. Dashed lines indicate different relations between $\tau_i$ and $\tau_{ev}$. $d_d^0 = 10$ μm for (a) and (c), $\phi_l = 1.0$ for (b) and (d). $H_0 = 25$ km and $M_0 = 9$.

The diagrams of ignition delay versus droplet evaporation time under various liquid *n*-heptane ERs and initial droplet diameters are shown in Fig. 14. Here $H_0 = 25$ km and $M_0 = 9$. Two background gas ERs are considered, i.e., $\phi_g = 0.4$ (Figs. 14a-14b) and 1.0 (Figs. 14c-14d). $d_d^0$ is 10 μm in Figs. 14(a) and 14(c), whilst $\phi_l$ is 1.0 in Figs. 14(b) and 14(d). In genreal, the ignition and evaporation process is longer in $\phi_g = 1.0$ than in $\phi_g = 0.4$. This is because the gas temperature behind the OSW (i.e., initial temperature in the 0D caluclations) is higher under fuel-lean conditions [13].

It is found from Fig. 14 that as the liquid ER increases or the droplet diameter decreases, the ignition delay time increases. This is because the higher (faster) energy absorption by droplet phase change from heavier droplet loading (smaller diameter). This leads to weakened chemical reactivity of the gas mixture. Moreover, the evaporation time is longer with increased liquid ER or droplet diameter. This is because the gas temperature in the reactor decreases with the evaporation heat transfer when the liquid ER is high. Also, the smaller the specific surface area for larger droplets, the smaller the heat transfer area between droplets and gas, resulting in slower evaporation. For $d_d^0$ = 10 μm in Figs 14(a) and 14(c), the ignition delay time is consistently longer than the evaporation time for all liquid ER's, which indicates that thermal runaway always occurs in a homogeneous mixture due to earlier completion of droplet evaproation. Conversely, in Figs. 14(b) and 14(d), the droplet diameter has significant effects on the mixture composition at the ignition point: heterogeneous ignition only occurs when the droplets are coarse. When the droplet diameter is smaller (i.e., $d_d^0$ < 16 μm for $\phi_g$ = 0.4 and $d_d^0$ < 22 μm for $\phi_g$ = 1.0), homogeneous ignition is observed. Under these conditions, the droplet evaporation occurs in the induction stage, and hence is mainly influenced by the properties of the shocked gas, rather than the combustion heat release.

### 5.5 $\tau_i - \tau_e$ diagram and its implication for two-phase ODW initiation

It is already shown in Fig. 5 that the ignition delay and chemical excitation time can well identify the oblique detonation transition mode. Therefore, Figs. 15 and 16 further compile them from the foregoing 0D two-phase calculations, with background gas ERs of $\phi_g$ = 0.4 and 1.0, respectively. The flight altitude $H_0$ increases from 21 to 43 km, as indicated by the arrows in Figs. 15(a) and 16(a). The implications for the symbols are given in Table 3.

One can see from Fig. 15(a) that, when $M_0 \leq 10$ (black/red/blue symbols), with increased $H_0$, the ignition delay time $\tau_i$ increases first and then decreases, consistent with the results in Figs. 13(a) and 13(c). In these cases, the autoignition process is sensitive to the change of post-OSW thermodynamic state, as discussed in Fig. 13. Nonetheless, when $M_0 > 10$ (green/magenta symbols), $\tau_i$ increases with $H_0$. In these cases, change of the inflow liquid ER has limited effects on the ignition, which is mainly controlled by the post-OSW thermodynamic state instead. Besides, the chemical excitation time $\tau_e$ generally increases (hence slower thermal runaway) with $H_0$ due to quickly reduced gas pressure, whilst $\tau_e$ generally decreases with $M_0$ (when other parameter are kept constant) due to increased gas temperature. Similar observations can also be made for other droplet diameters (i.e., 10 and 50 μm) in Figs. 15(b) and 15(c).

Moreover, addition of the *n*-heptane droplets would lower the gas temperature and increase the *n*-heptane vapor mass fraction. Therefore, the ignition timescales (i.e., $\tau_e$ and $\tau_i$) would be affected. As can be found in Fig. 15(a), when $M_0 \leq 10$, $\tau_i$ increases with liquid ER $\phi_l$ (from square to diamond), and this trend becomes relatively weak as $d_d^0$ increases in Figs. 15(b) and 15(c). This is because the change of gas thermochemical state is small when coarser droplets are loaded. Moreover, with increased $\phi_l$, the evaporation of a large number of droplets leads to high energy absorption. The resultant gas temperature reduction weakens the reactions and therefore $\tau_e$ may increase. When $M_0 >$ 10, both timescales are less affected by $\phi_l$ or $d_d^0$. The reason is that as $M_0$ is beyond 10, both temperature and pressure behind the OSW are sufficeintly high, and the influences of heat and mass transfer from the fuel droplets on the gas thermochemical state are of secondary importance.

We further examine how the ratio of the two timescales is affected by the liquid fuel properties and flight conditions, and their inplications for spray ODW initiation. In Figs. 15(a)-15(c), when $M_0 \leq 10$, $\tau_e/\tau_i$ is generally below 0.2 (fast thermal runaway) and increases with flight altitude $H_0$ or Mach number $M_0$. Based on the discussion in Fig. 5, smooth transition from OSW to ODW is more likely to occur when $H_0$ and/or $M_0$ is higher. Moreover, $\tau_e/\tau_i$ increases as the liquid ER decreases. This implies that addition of the liquid fuel droplets makes the ODW more likely to be initiated abruptly.

When $M_0 > 10$, the ratio of $\tau_e$ to $\tau_i$ is consistently high, regardless of the fuel and inflow conditions ($\phi_l$, $d_d^0$, $H_0$ and $M_0$), as demonstrated in Fig. 15. Therefore, we can conjecture that the liquid *n*-$C_7H_{16}$ ODW is mainly initiated smoothly from the OSW at high $M_0$. Also, the fuel spray properties exhibit negligible influences on the varation of transition mode.

Table 3. Impications of the symbols in Figs. 15 and 16.

| Symbol | | Parameter |
|---|---|---|
| Symbol colour for flight Mach number $M_0$ | Black | $M_0 = 8$ |
| | Red | $M_0 = 9$ |
| | Blue | $M_0 = 10$ |
| | Green | $M_0 = 11$ |
| | Magenta | $M_0 = 15$ |
| Symbol shape for liquid ER $\phi_l$ | Square | $\phi_l = 0.0$ (droplet-free) |
| | Cricle | $\phi_l = 0.4$ |
| | Triangle | $\phi_l = 1.0$ |
| | Diamond | $\phi_l = 2.0$ |

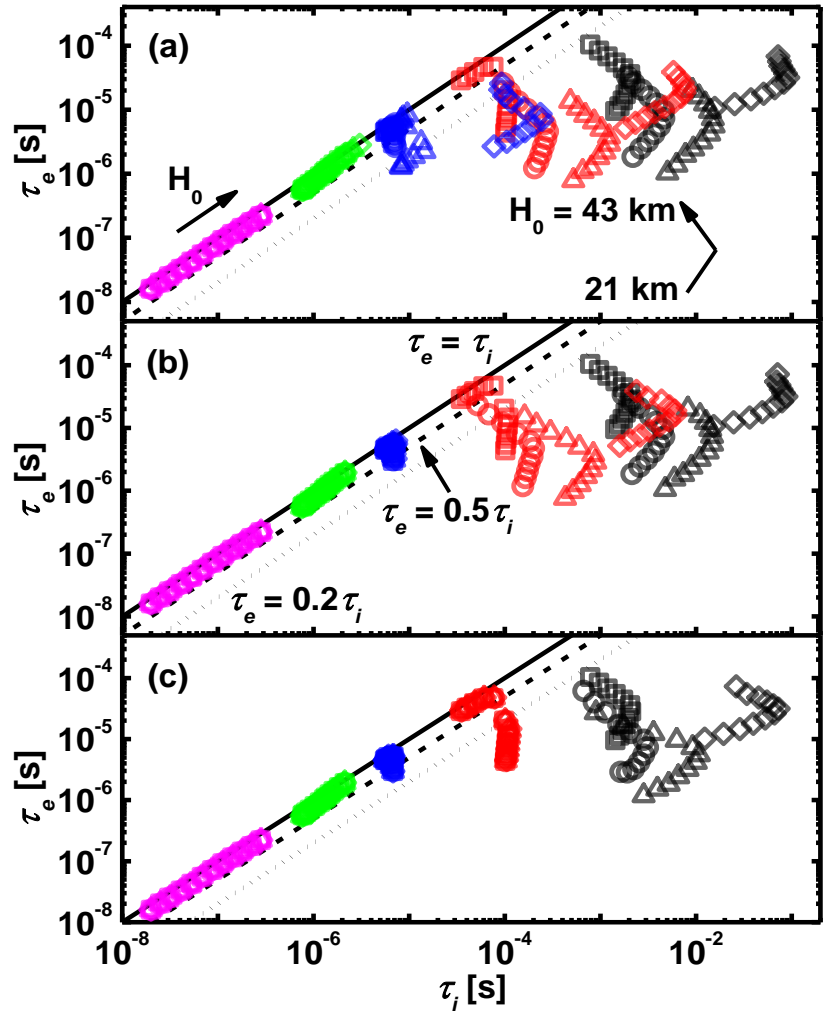

Figure 15: Chemical excitation time versus ignition delay time with initial droplet diameters of (a) 5, (b) 10, and (c) 50 μm. $H_0 = 21 – 43$ km and $\phi_g = 0.4$. Symbol legend in Table 3.

By comparing Figs. 15 and 16 (with different gas ERs, 0.4 and 1.0, respectively), one can have the qualitatively similar observations. Nonetheless, when the background gas ER is unity in Fig. 16, the ignition delay time and chemical excitation time are generally longer, and more likely to form abrupt transition mode. This is consistent with the results in previous studies [13]. Besides, the Mach number range within which $\tau_e/\tau_i$ is significantly affected by flight altitude and/or liquid fuel properties is slightly expanded, from $M_0$ = 8-10 in $\phi_g$ = 0.4 to $M_0$ = 8-11 in $\phi_g$ = 1.0.

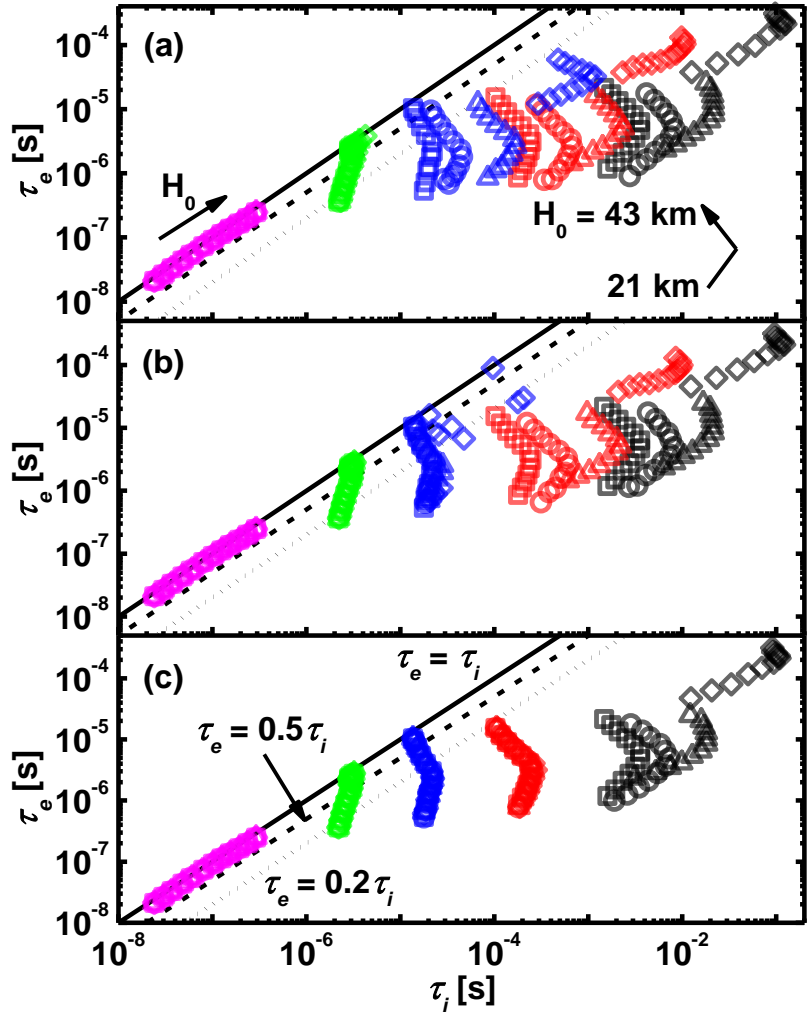

Figure 16: Chemical excitation time versus ignition delay time with initial droplet diameters of (a) 5, (b) 10, and (c) 50 μm. $H_0$ = 21 – 43 km and $\phi_g$ = 1.0. Symbol legend in Table 3.

The reader should be reminded that the above analysis based on Figs. 15 and 16 is based on fuel chemical kinetics under ODW-relevant flight conditions. However, due to multiple possible factors existing in spray ODW initiation, e.g., droplet fragmentation and breakup, phase change, polydispersity, and boundary layer, high-fidelity multi-dimensional simulations with detailed fuel chemistry should be conducted in our future work to further determine the accurate chemical timescale ratio for transition mode prediction and reveal the detailed transition dynamics.

## 6. Conclusions

Autoignition of *n*-heptane droplet/vapor/air mixtures behind an oblique shock wave are computationally studied, through Eulerian-Lagrangian method and a skeletal chemical mechanism. The effects of inflow gas/liquid equivalence ratio, droplet diameter, flight altitude, and Mach number are studied. The chemical timescales and transient of the two-phase gas autoignition are discussed in detail. The following conclusions can be drawn:

(1) The ratio of chemical excitation time to ignition delay time can reasonably identify the shock-to-detonation transition mode based on extensive previous studies. When the ratio is relatively high (e.g., higher than 0.5 for *n*-$C_7H_{16}$), the combustion heat release is slow and smooth transition is more likely to occur.

(2) Homogeneous and heterogeneous ignitions are observed, depending on the composition of the mixture when ignition occurs. If the droplets complete the evaporation before (after) ignition, homogeneous (heterogeneous) ignition can be found. In heterogeneous ignition, direct interactions between the evaporating droplets and induction/ignition process exist. Continuous reactions are also observed when the droplets continue evaporating in the post-ignition stage. Through chemical explosive mode analysis, alternate changes of the chemical explosive propensity after the ignition are revealed.

(3) The evaporating droplets absorb energy from the gas phase and transfers mass to the gas phase. The heat transfer from evaporating droplets significantly delays the autoignition of *n*-heptane vapor. Also, addition of fuel droplets to the fuel-lean mixture is found to promote the ignition of the gas phase.

(4) In the two-phase *n*-heptane mixture autoignition process, the ignition delay time decreases exponentially with flight Mach number, and increases first and then decreases with the flight altitude. As the liquid ER increases, both ignition delay time and droplet evaporation time increase. With increased droplet diameter, the ignition delay time decreases and the evaporation time increases. In low flight Mach number, the ignition delay time is longer than the droplet evaporation time.

(5) For relatively low Mach numbers (< 10), the ratio of the chemical excitation time to ignition delay time generally increases with the flight altitude or Mach number. It increases as the liquid ER decreases or droplet diameter increases. While for $M_0 > 10$, the ratio changes marginally with fuel and inflow conditions. The oblique detonation wave is more likely to be initiated with a smooth mode at high altitude or Mach number. Moreover, it is more probable to have an abrupt transition with fine fuel droplets.

## Acknowledgements

This work used the computational resources of the National Supercomputing Centre, Singapore (https://www.nscc.sg/). HG is supported by the China Scholarship Council (No. 202006680013). Professor Zhuyin Ren at Tsinghua University is thanked for sharing the CEMA routines.